\documentclass{pasj00}
\begin{document}
\SetRunningHead{J.\ Fukue}
{Radiative Flow in a Luminous Disk II}
\Received{2005/06/18}
\Accepted{2005/11/08}

\title{Radiative Flow in a Luminous Disk II}

\author{Jun \textsc{Fukue}} 
\affil{Astronomical Institute, Osaka Kyoiku University, 
Asahigaoka, Kashiwara, Osaka 582-8582}
\email{fukue@cc.osaka-kyoiku.ac.jp}


\KeyWords{
accretion, accretion disks ---
astrophysical jets ---
galaxies: active ---
radiative transfer ---
relativity ---
X-rays: stars
} 

\maketitle


\begin{abstract}
Radiatively-driven transfer flow perpendicular to a luminous disk
is examined in the subrelativistic regime of $(v/c)^1$,
taking into account the gravity of the central object.
The flow is assumed to be vertical, and
the gas pressure is ignored,
while internal heating is assumed to be proportional to the gas density.
The basic equations were numerically solved
as a function of the optical depth,
and the flow velocity, the height, the radiative flux, and
the radiation pressure were obtained
for a given radius, an initial optical depth,
and initial conditions at the flow base (disk ``inside''),
whereas the mass-loss rate was determined as an eigenvalue
of the boundary condition at the flow top (disk ``surface'').
For sufficiently luminous cases,
the flow resembles the case without gravity.
For less-luminous cases, however,
the flow velocity decreases, and
the flow would be impossible due to the existence of gravity
in the case that the radiative flux is sufficiently small.
Application to a supercritical accretion disk with mass loss
is briefly discussed.
\end{abstract}

\section{Introduction}

Mass outflow from a luminous disk
is a clue to the formation mechanism of
astrophysical jets and winds in the active objects.
In particular, in a supercritical accretion disk,
the mass-accretion rate highly exceeds the critical rate, 
the disk local luminosity exceeds the Eddington one,
and mass loss from a disk surface 
driven by radiation pressure would take place
(see Kato et al. 1998 for a review of accretion disks).

So far, radiatively driven outflow from a luminous disk
has been extensively studied by many researchers
(Bisnovatyi-Kogan, Blinnikov 1977; Katz 1980; Icke 1980; Melia, K\"onigl 1989; 
Misra, Melia 1993; Tajima, Fukue 1996, 1998; Watarai, Fukue 1999;
Hirai, Fukue 2001; Fukue et al. 2001; Orihara, Fukue 2003),
as on-axis jets (Icke 1989; Sikora et al. 1996; Renaud, Henri 1998;
Luo, Protheroe 1999; Fukue 2005a),
as outflows confined by a gaseous torus
(Lynden-Bell 1978; Davidson, McCray 1980; Sikora, Wilson 1981; Fukue 1982),
or as jets confined by the outer flow or corona
(Sol et al. 1989; Marcowith et al. 1995; Fukue 1999),
and by numerical simulations
(Eggum et al. 1985, 1988).
Line-driven outflows were also extensively examined
(Drew, Verbunt 1985; Vitello, Shlosman 1988;
Proga et al. 1998; Murray, Chiang 1995; Proga et al. 2000;
Proga, Kallman 2004).
In almost all of these studies, however,
the luminous disk was treated as an external radiation source,
and the radiation transfer in the flow was not solved.

Radiation transfer in the disk, on the other hand,
was investigated in relation to the structure
of a static disk atmosphere and
the spectral energy density from the disk surface
(e.g., Meyer, Meyer-Hofmeister 1982; Cannizzo, Wheeler 1984;
Shaviv, Wehrse 1986; Adam et al. 1988;
Hubeny 1990; Ross et al. 1992; Artemova et al. 1996;
Hubeny, Hubeny 1997, 1998; Hubeny et al. 2000, 2001;
Davis et al. 2005; Hui et al. 2005;
see also Mineshige, Wood 1990).
In these studies, however,
the vertical movement and the mass loss were not considered.

Recently,
mass outflow as well as radiation transfer have been examined
for the first time in the subrelativistic and
fully relativistic cases (Fukue 2005b, c).
In these studies,
the gravity of the central object and the gas pressure
were both ignored as a first step.
In this paper,
we thus consider the radiatively driven vertical outflow
-- {\it moving photosphere} -- in a luminous flat disk
within the framework of radiation transfer
in the subrelativistic regime of $(v/c)^1$,
while taking into account the gravity of the central object,
although the gas pressure is ignored.

In the next section
we describe the basic equations in the vertical direction.
In section 3
we show our numerical examination of the radiative flow.
In section 4
we briefly apply the present model
to the case of a supercritial accretion disk.
The final section is devoted to concluding remarks.


\section{Basic Equations}

Let us suppose a luminous flat disk, inside of which
gravitational or nuclear energy is released
via viscous heating or other processes.
The radiation energy is transported in the vertical direction,
and the disk gas, itself, also moves in the vertical direction
due to the action of radiation pressure.
For simplicity, in the present paper,
the radiation field is considered to be sufficiently intense that
the gas pressure can be ignored (cf. Fukue 2005b):
tenuous cold normal plasmas in the super-Eddington disk,
cold pair plasmas in the sub-Eddington disk, or
dusty plasmas in the sub-Eddington disk.
As for the order of the flow velocity $v$,
we consider the subrelativistic regime,
where terms of the first order of $(v/c)$ are retained,
in order to take into account radiation drag.
Under these assumptions,
the radiation hydrodynamic equations
for steady vertical ($z$) flows without rotation are described as follows
(Hsieh, Spiegel 1976; Kato et al. 1998
for the full set of basic equations).

The continuity equation is
\begin{equation}
   \rho v = J ~(={\rm const}),
\label{rho1}
\end{equation}
where $\rho$ is the gas density, $v$ the vertical velocity, and
$J$ the mass-loss rate per unit area.
The equation of motion is
\begin{equation}
   v\frac{dv}{dz} = -\frac{GMz}{(R-r_{\rm g})^2 R}
                    +\frac{\kappa_{\rm abs}+\kappa_{\rm sca}}{c}
                    \left[ F - (E+P)v \right],
\label{v1}
\end{equation}
where $M$ is the mass of the central object,
$R$ $=\sqrt{r^2+z^2}$, $r$ being the radius;
$r_{\rm g}$ ($=2GM/c^2$) is the Schwarzschild radius,
$\kappa_{\rm abs}$ and $\kappa_{\rm sca}$
are the absorption and scattering opacities (gray),
$E$ is the radiation energy density, $F$ is the radiative flux, and
$P$ is the radiation pressure,
measured in the inertial frame, respectively.
As for the gravity,
we adopt the pseudo-Newtonian potential (Paczy\'nski, Wiita 1980).

When the gas pressure is ignored,
the advection terms of the energy equation 
are dropped (cf. Kato et al. 1998), 
and heating is balanced with the cooling,
\begin{equation}
   0 = q^+ - \rho \left( j - c\kappa_{\rm abs} E
                  + \kappa_{\rm abs} \frac{2Fv}{c} \right),
\label{j1}
\end{equation}
where $q^+$ is the internal heating and $j$ is the emissivity.
The last term in the parentheses is the work done by the radiation field
under the present approximation.

Under the assumption of steady one-dimensional flow,
and under the present approximation,
the radiation field equations (Kato et al. 1998) become
\begin{equation}
   \frac{dF}{dz} = \rho
         \left[ j - c\kappa_{\rm abs} E
         + (\kappa_{\rm abs}-\kappa_{\rm sca}) \frac{Fv}{c} \right],
\label{F1}
\end{equation}
\begin{equation}
   \frac{dP}{dz} = \frac{\rho v}{c^2} \left( j - c\kappa_{\rm abs} E \right)
           -\rho \frac{\kappa_{\rm abs}+\kappa_{\rm sca}}{c}
                    \left[ F - (E+P)v \right],
\label{P1}
\end{equation}
and the closure relation becomes
\begin{equation}
   P = \frac{1}{3}E + \frac{4}{3}\frac{Fv}{c^2}.
\label{E1}
\end{equation}

Eliminating $j$ and $E$ with the help of equations (\ref{j1}) and (\ref{E1}),
and introducing the optical depth by
\begin{equation}
    d\tau = - ( \kappa_{\rm abs}+\kappa_{\rm sca} ) \rho dz,
\end{equation}
we can rearrange the basic equations up to the order of $(v/c)^1$ as
\begin{eqnarray}
   cJ\frac{dv}{d\tau} &=& \frac{c}{\kappa_{\rm abs}+\kappa_{\rm sca}}
                          \frac{GMz}{(R-r_{\rm g})^2 R}
                          -\left( F - 4Pv \right),
\label{v}
\\
   J\frac{dz}{d\tau} &=& -\frac{1}{\kappa_{\rm abs}+\kappa_{\rm sca}}v,
\label{z}
\\
   \frac{dF}{d\tau} &=& -\frac{ q^+ }
                              { (\kappa_{\rm abs}+\kappa_{\rm sca}) \rho }
                        +F\frac{v}{c},
\label{F}
\\
   c\frac{dP}{d\tau} &=& F - 4Pv
                         -\frac{ q^+ }
                { (\kappa_{\rm abs}+\kappa_{\rm sca}) \rho }
                         \frac{v}{c}.
\label{P}
\end{eqnarray}

In this paper 
we assume that the heating $q^+$ is proportional to the gas density $\rho$
for simplicity (cf. Fukue 2005b), and set the following relation:
\begin{equation}
   \frac{ q^+ } { (\kappa_{\rm abs}+\kappa_{\rm sca}) \rho }
     = \frac{F_0}{\tau_0},
\end{equation}
where $F_0$ is a parameter
that determines the magnitude of heating,
and therefore the magnitude of the radiation field.

We solved equations (\ref{v})--(\ref{P})
for appropriate boundary conditions.

As for the boundary conditions,
we imposed the following cases.
At the flow base (disk ``inside'')
with radius $r$ and an arbitrary optical depth $\tau_0$,
the flow velocity is zero ($v=0$),
the height is zero ($z=0$),
the radiative flux is zero ($F=0$), and
the radiation pressure is $P_0$,
where subscript 0 denotes the values at the flow base.
At the flow top (disk ``surface''),
where the optical depth is zero,
the radiation fields should approximately have the values
above the luminous infinite disk, and therefore
satisfy the condition  $cP_{\rm s}/F_{\rm s}=2/3$,
where subscript s denotes the values at the flow top (Fukue 2005b).
Hence, the parameters are
$r$, $\tau_0$, $F_0$, and $P_0$ at the flow base,
and the mass-loss rate $J$ is determined
as an eigenvalue of the boundary condition at the flow top.

In numerical calculations,
we start from the asymptotic solutions near to $\tau_0$:
\begin{eqnarray}
   t &=& \tau_0 - \tau,   ~~~t \ll\tau_0,
\label{IC1}
\\
   v &\sim& \frac{1}{2cJ} \frac{F_0}{\tau_0} t^2,
\label{IC2}
\\
   z &\sim& \frac{1}{6(\kappa_{\rm abs}+\kappa_{\rm sca})cJ^2}
          \frac{F_0}{\tau_0} t^3,
\label{IC3}
\\
   F &\sim& \frac{F_0}{\tau_0} t,
\label{IC4}
\\
   P &\sim& P_0 - \frac{1}{2c} \frac{F_0}{\tau_0} t^2,
\label{IC5}
\end{eqnarray}
and solve equations (\ref{v})--(\ref{P})
down to the optical depth to $\tau=0$.
We then obtain the final values of 
$v_{\rm s}$, $z_{\rm s}$, $F_{\rm s}$, and $P_{\rm s}$ at the flow top.
Generally, the final values do not satisfy the boundary condition of
$cP_{\rm s}/F_{\rm s}=2/3$.
We thus iterate the value of $J$ so as to satisfy the boundary condition.
Technically, in some cases,
we iterate $P_0$ instead of $J$.

Finally, we normalize the physical quantities
in terms of the speed of light $c$, the Schwarzschild radius $r_{\rm g}$, 
and the Eddington luminosity $L_{\rm E}$
[$=4\pi cGM/(\kappa_{\rm abs}+\kappa_{\rm sca})$],
since we consider the luminous disk around a black hole.
That is,
the normalized quantities are
$\hat{v}=v/c$, $\hat{z}=z/r_{\rm g}$,
$\hat{F}=F/L_{\rm E}/(4\pi r_{\rm g}^2)$,
$\hat{P}=cP/L_{\rm E}/(4\pi r_{\rm g}^2)$,
and $\hat{J}=c^2J/L_{\rm E}/(4\pi r_{\rm g}^2)$.
Moreover, the basic equations (\ref{v})--(\ref{P})
are expressed as
\begin{eqnarray}
   J\frac{dv}{d\tau} &=&  \frac{z}{(R-1)^2 R}
                          -\left( F - 4Pv \right),
\label{v_nor}
\\
   J\frac{dz}{d\tau} &=& -2v,
\label{z_nor}
\\
   \frac{dF}{d\tau} &=& -\frac{ F_0 } { \tau_0 }  +Fv,
\label{F_nor}
\\
   \frac{dP}{d\tau} &=& F - 4Pv - \frac{F_0}{\tau_0}v,
\label{P_nor}
\end{eqnarray}
where the symbol ``hat'' (say, $\hat{v}$) is dropped.

\section{Radiative Flow under Gravity}

In this section
we show the radiative vertical flow in the luminous disk
under the influence of the gravity of the central object.
In order to obtain the solution,
as already stated,
we numerically solve equations (\ref{v_nor})--(\ref{P_nor}),
starting from asymptotic solutions near to $\tau=\tau_0$
with appropriate initial conditions,
down to $\tau=0$ so as to satisfy
appropriate boundary conditions there.
The parameters are
the initial radius $r$ on the disk,
the initial optical depth $\tau_0$,
which relates to the disk surface density,
the heating rate coefficient $F_0$,
which roughly expresses the radiative flux at the flow top and
is the measure of the strength
of radiation field to gravity,
and the initial radiation pressure $P_0$ at the disk base,
which connects with the radiation pressure gradient
in the vertical direction and
relates to the disk internal structure.
The mass-loss rate $J$ is determined
as an eigenvalue of the boundary condition at the flow top.

Several examples of numerical calculations
are shown in figures 1 and 2.

\begin{figure}
  \begin{center}
  \FigureFile(80mm,80mm){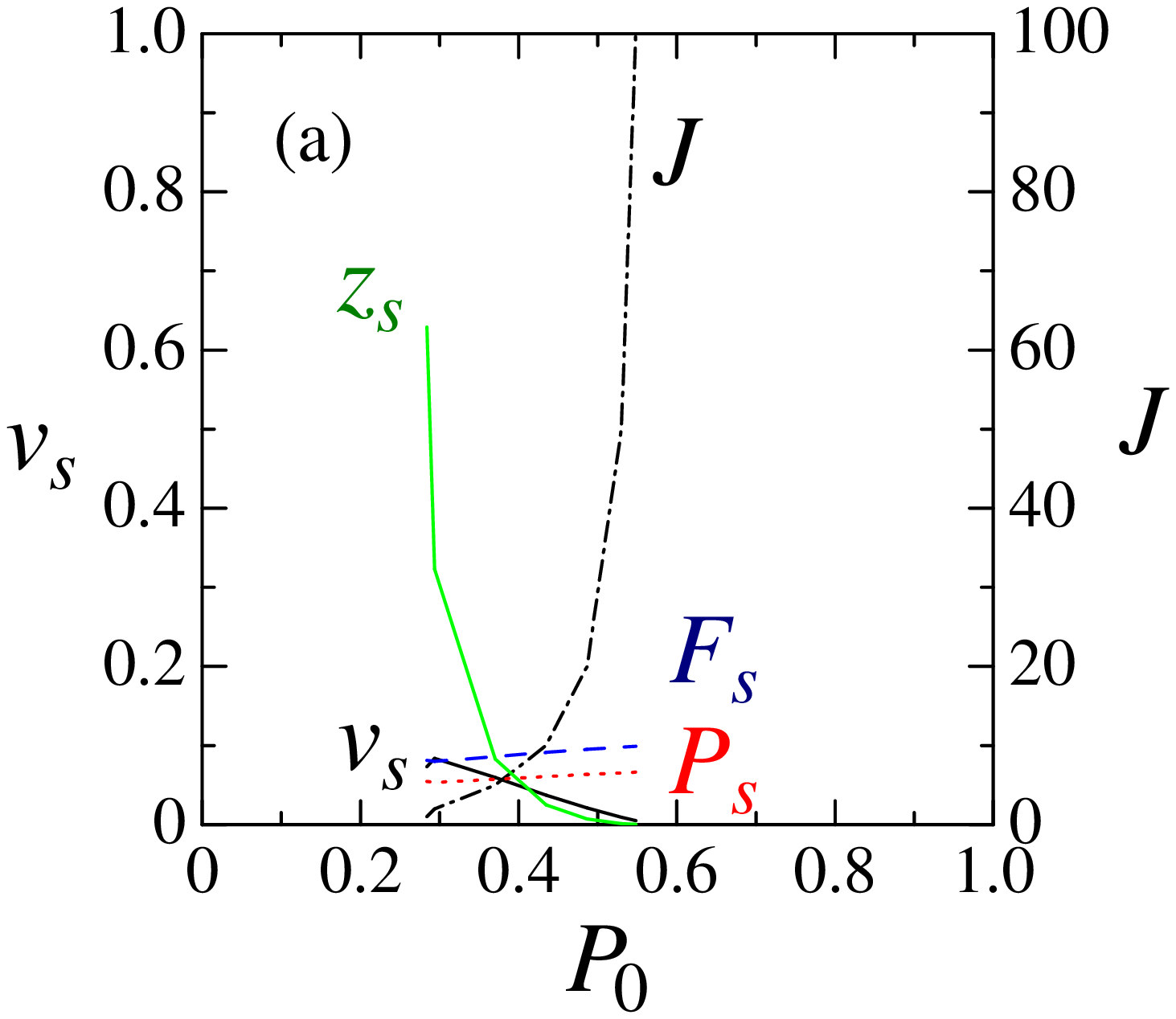}
  \end{center}
  \begin{center}
  \FigureFile(80mm,80mm){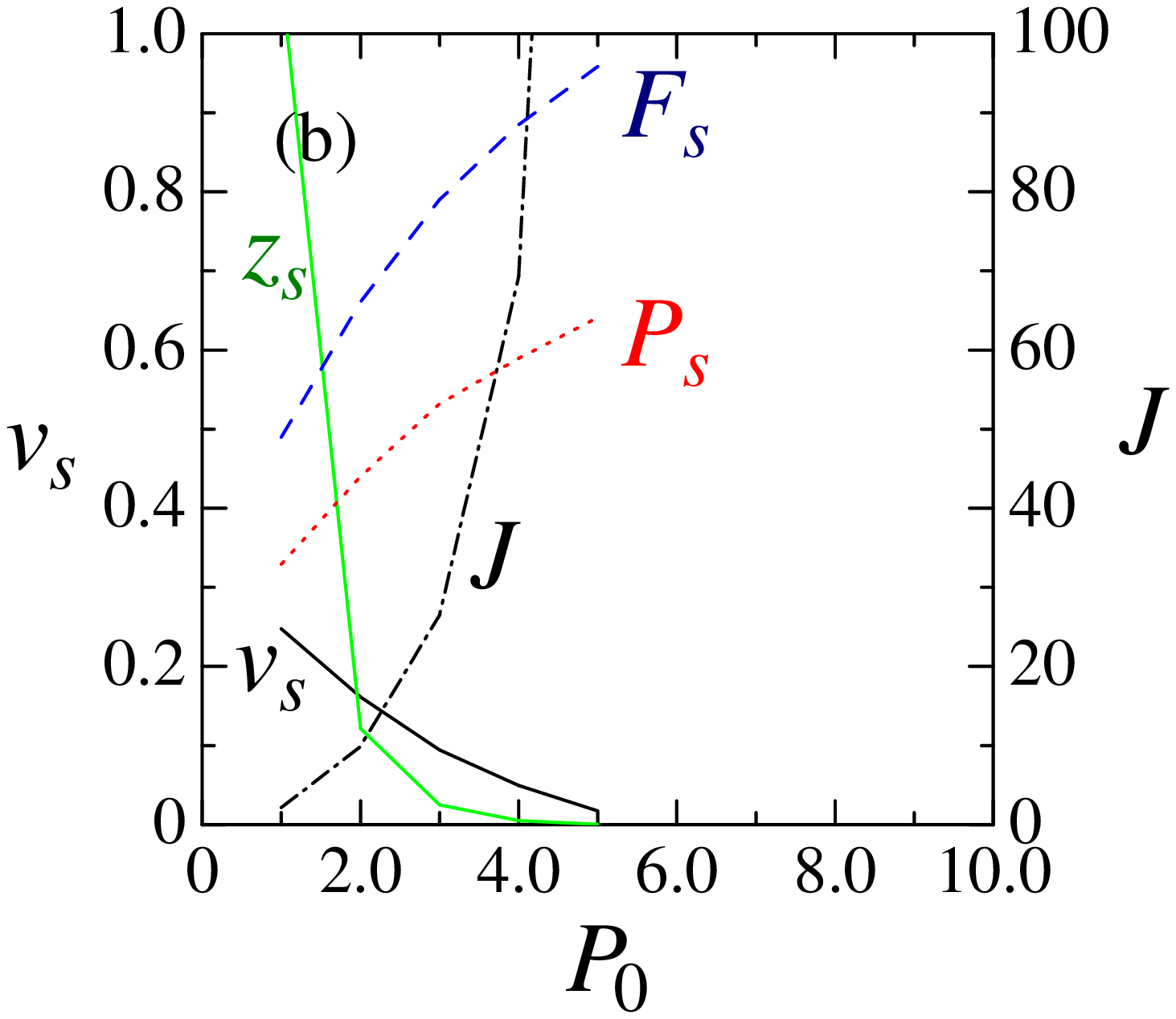}
  \end{center}
\caption{
Final velocity $v_{\rm s}$ (solid curves),
flow height $z_{\rm s}$ (solid ones),
radiative flux $F_{\rm s}$ (dashed ones),
radiation pressure $P_{\rm s}$ (dotted ones)
at the flow top,
and the mass-loss rate $J$ (chain-dotted ones),
as a function of $P_0$ at the flow base.
The parameters are
$r=3$, $\tau_0=10$, and (a) $F_0=0.1$ and (b) $F_0=1$.
The quantities are normalized in units of $c$, $r_{\rm g}$, and 
$L_{\rm E}/(4\pi r_{\rm g}^2)$.
The leftside scale of the ordinates is
for quantities except for $J$, while
the rightside is for $J$.
}
\end{figure}

\begin{figure}
  \begin{center}
  \FigureFile(80mm,80mm){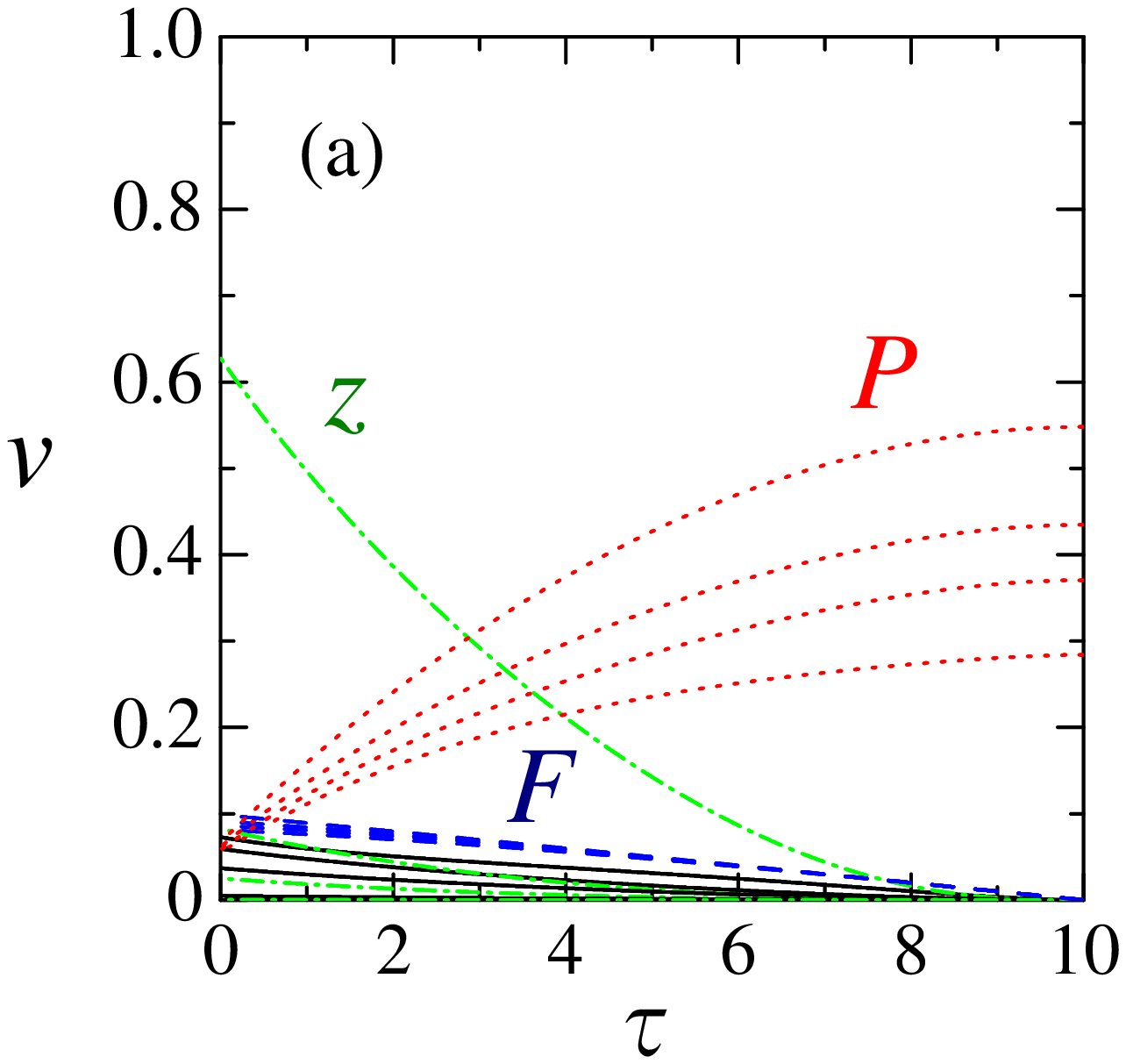}
  \end{center}
  \begin{center}
  \FigureFile(80mm,80mm){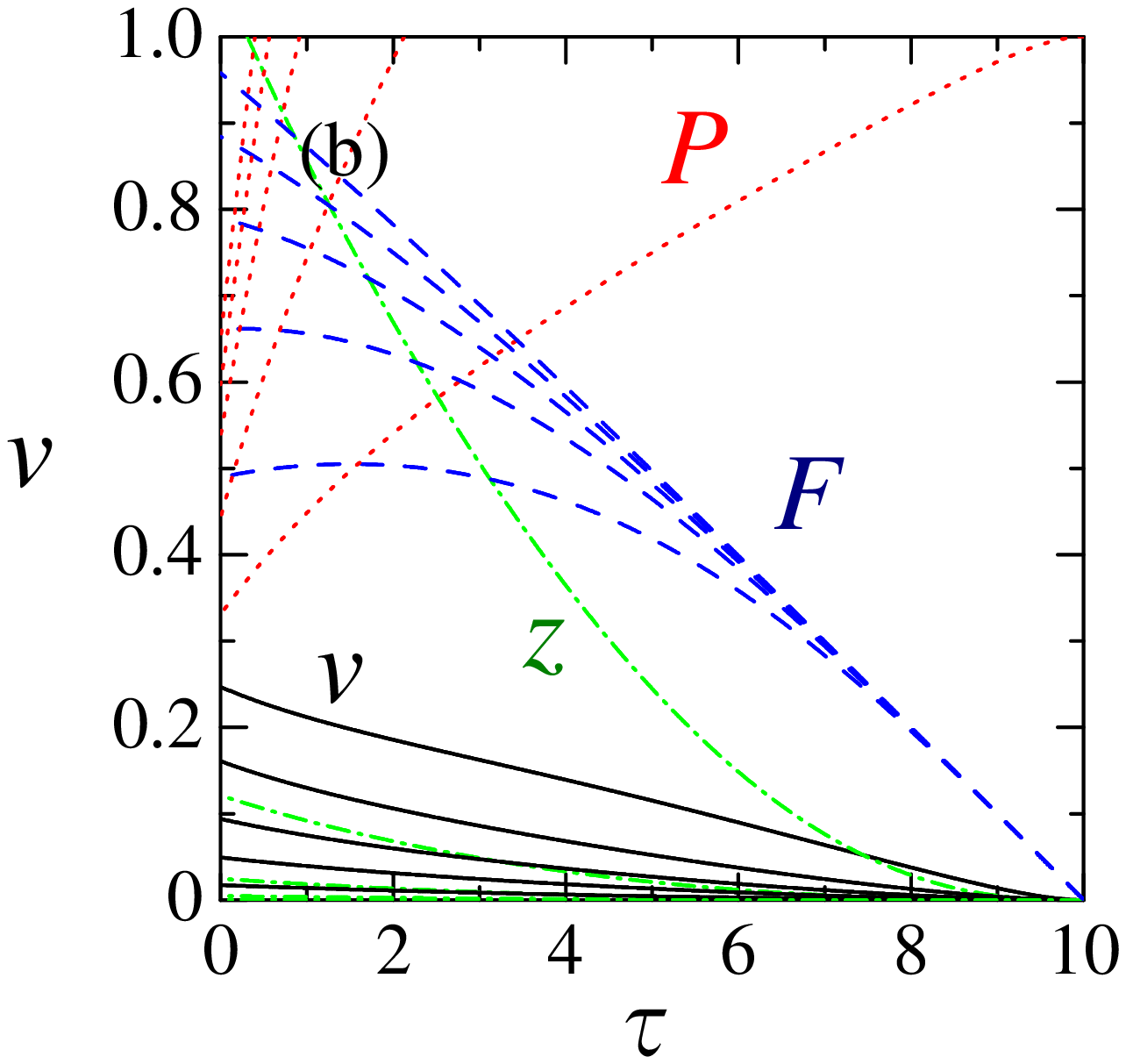}
  \end{center}
\caption{
Flow velocity $v$ (solid curves),
flow height $z$ (chain-dotted ones), 
radiative flux $F$ (dashed ones),
and radiation pressure $P$ (dotted ones),
as a function of the optical depth $\tau$.
The parameters are $r=3$, $\tau_0=10$, and
(a) $F_0=0.1$ and (b) $F_0=1$.
In addition, the values of $P_0$ are
(a) 0.2840, 0.3705, 0.4346, and 0.5480 and
(b) 1, 2, 3, 4, and 5
 from top to bottom of $v$ and $z$ and
 from bottom to top of $F$ and $P$.
The quantities are normalized in units of $c$, $r_{\rm g}$, and 
$L_{\rm E}/(4\pi r_{\rm g}^2)$.
}
\end{figure}

In figure 1
we show 
the final velocity $v_{\rm s}$ (solid curves),
the flow height $z_{\rm s}$ (solid ones),
the radiative flux $F_{\rm s}$ (dashed ones),
the radiation pressure $P_{\rm s}$ (dotted ones)
at the flow top,
and the mass-loss rate $J$ (chain-dotted ones),
as a function of $P_0$ at the flow base
for $r=3$, $\tau_0=10$, and several values of $F_0$.
The quantities are normalized in units of $c$, $r_{\rm g}$, and 
$L_{\rm E}/(4\pi r_{\rm g}^2)$.

As can be seen in figure 1,
the mass-loss rate increases as the initial radiation pressure increases,
while the flow final speed increases
as the initial radiation pressure and the loaded mass decrease.
These properties are similar to those
in a previous study without gravity (Fukue 2005b).
In particular,
when the disk luminosity is high (figure 1b),
the results are quite similar to the previous case without gravity.
In the case where the disk luminosity is low (figure 1a),
on the other hand,
the final velocity becomes small, as easily expected.
Furthermore,
if the disk luminosity is sufficiently small,
the radiative flow becomes impossible
due to the effect of gravity.

In addition, as can be seen in figure 1, 
in order for the flow to exist,
the radiation pressure $P_0$ at the flow base is restricted
in some range similar to the previous case.
That is, in the previous study without gravity (Fukue 2005b),
the initial pressure $P_0$ is proved to be restricted
in the range of $2/3 < cP_0/F_{\rm s} < 2/3 + \tau_0/2$
in the case where heating is included.
In the present case,
when the disk luminosity (characterized by $F_0$) is high (figure 1b),
$P_0$ is restricted in the similar range.
When the disk luminosity is low (figure 1a), on the other hand,
the values of $P_0$ become small, but
is yet restricted in some range.
At the upper limit of these ranges,
the loaded mass diverges and 
both the flow final speed and flow height become zero.
On the other hand, at the lower limit,
the loaded mass becomes zero and
the flow final speed becomes the maximum.

In figure 2
we show the flow velocity $v$ (solid curves),
the flow height $z$ (chain-dotted ones), 
the radiative flux $F$ (dashed ones),
and the radiation pressure $P$ (dotted ones),
as a function of the optical depth $\tau$
for $r=3$, $\tau_0=10$, and several values of $P_0$
in a few cases of $F_0$.
The quantities are normalized in units of $c$, $r_{\rm g}$, and 
$L_{\rm E}/(4\pi r_{\rm g}^2)$.

When the initial radiation pressure $P_0$ at the flow base is large,
the pressure gradient between the flow base and the flow top is
also large.
As a result, the loaded mass $J$ also becomes large,
but the flow speed $v$ is small.
When the initial radiation pressure $P_0$ is small,
on the other hand,
the pressure gradient becomes small, and
the loaded mass also is small, but
the flow speed becomes large.

When the disk luminosity (characterized by $F_0$) is large (figure 2b),
the flow velocity as well as the radiation fields become large.
When the disk luminosity is small (figure 2a), on the other hand,
the flow velocity and the radiation fields are small.

In the limit of small $F_0$ (see figure 2a),
the radiation fields approach the non-relativistic case
of $v \ll c$.
In such a case,
equations (\ref{F_nor}) and (\ref{P_nor})
are analytically integrated to give
\begin{eqnarray}
   F &=& \frac{F_{\rm s}}{\tau_0} (\tau_0 - \tau),
\\
   cP &=& cP_0 - \frac{1}{2}\frac{F_{\rm s}}{\tau_0} (\tau - \tau_0)^2,
\end{eqnarray}
where $F_{\rm s}=F_0$ in this limit.
In this limit, the relation derived in the previous study (Fukue 2005b)
again holds:
\begin{equation}
   \frac{cP_0}{F_{\rm s}} = \frac{2}{3} + \frac{1}{2}\tau_0.
\end{equation}
Hence,
\begin{equation}
   T_{\rm c}^4 = \frac{3}{4} T_{\rm eff}^4
              \left( \frac{2}{3} + \frac{1}{2}\tau_0 \right),
\end{equation}
where $T_{\rm c}$ is the temperature at the flow base
(i.e., $P_0=aT_{\rm c}^4/3$), and
$T_{\rm eff}$ the effective temperature at the flow top
(i.e., $F_{\rm s}=\sigma T_{\rm eff}^4$).
This is just the modified Milne approximation.


\section{Radiative Flow in the Supercritical Disk}

In this section
we apply the present model to
the mass outflow in the luminous supercritical accretion disks.

When the mass-accretion rate $\dot{M}$ in the disk
around a central object of mass $M$ highly exceeds
the critical rate $\dot{M}_{\rm crit}$,
defined by $\dot{M}_{\rm crit} \equiv L_{\rm E}/c^2$,
the disk is believed to be in the supercritial regime,
and the disk luminosity exceeds the Eddington one.
Such a supercritical accretion disk, a so-called slim disk,
has been extensively studied, both numerically and analytically
(Abramowicz et al. 1988; Eggum et al. 1988;
Szuszkiewicz et al. 1996; Beloborodov 1998;
Watarai, Fukue 1999; Watarai et al. 2000; Mineshige et al. 2000;
Fukue 2000; Kitabatake et al. 2002;
Ohsuga et al. 2002, 2003; Watarai, Mineshige 2003; Fukue 2004).
It was found that
the optically-thick supercritical disk is roughly expressed
by a self-similar model
(e.g., Watarai, Fukue 1999; Fukue 2000; Kitabatake et al. 2002;
Fukue 2004).
Except for the case of Fukue (2004), however,
many of these analytical models as well as numerical ones
did not consider the mass outflow from the disk surface.
Hence, in this paper
we adopt the model developted by Fukue (2004),
as a background supercritical disk model.

In the {\it critical accretion disk} constructed by Fukue (2004),
the mass-accretion rate is assumed to be regulated 
just at the critical rate with the help of wind mass-loss.
Outside some critical radius,
the disk is in a radiation-pressure dominated standard state,
while inside the critical radius
the disk is in a critical state,
where the excess mass is expelled by wind
and the accretion rate is kept to be just at the critical rate
at any radius.
Here, the critical radius is derived as
\begin{equation}
   r_{\rm cr} = \frac{9\sqrt{3}\sigma_{\rm T}}{16\pi m_{\rm p}c}
                \dot{M}_{\rm input}
              \sim 1.95 \dot{m} r_{\rm g},
\label{rcr}
\end{equation}
where $\dot{M}_{\rm input}$ is the accretion rate
at the outer edge of the disk, and
$\dot{m} = \dot{M}_{\rm input}/\dot{M}_{\rm crit}$.
Outside $r_{\rm cr}$, the accretion rate is constant,
while, inside $r_{\rm cr}$ the accretion rate would vary as
\begin{equation}
   \dot{M}(r) = \frac{16\pi cm_{\rm p}}{9\sqrt{3}\sigma_{\rm T}}r
   =\dot{M}_{\rm input} \frac{r}{r_{\rm cr}}.
\label{dotM}
\end{equation}

In such a critical accretion disk,
the disk thickness $H$ is conical as
\begin{equation}
   \frac{H}{r} = \sqrt{c_3} = \frac{1}{4}
                  \ln \left( 1 + \frac{\dot{m}}{20} \right),
\end{equation}
where $c_3$ is some numerical coefficient
determined by the similar procedure in Narayan and Yi (1994)
for optically-thin advection-dominated disks.
The second equality of this relation
comes from the numerical calculation (Watarai et al. 2000).
Although the mass loss was not considered in Watarai et al. (2000),
we adopted this relation as some measure:
when the normalized accretion rate $\dot{m}$ is 100,
the coefficient $\sqrt{c_3}$ becomes 0.448.

Furthermore, in Fukue (2004), several alternatives are discussed,
and some of them gives the physical quantities
of the critical accretion disk with mass loss as
\begin{eqnarray}
   \tau_0 &=& \frac{16\sqrt{6}}{\alpha} 
            \sqrt{\frac{r}{r_{\rm g}}},
\\
   F_{\rm s} &=& \sigma T_{\rm eff}^4 = \frac{3}{4}\sqrt{c_3}
                 \frac{L_{\rm E}}{4\pi r^2},
\\
   cP_0 &=& \frac{cGM}{\kappa}\sqrt{c_3}\tau \frac{1}{r^2},
\end{eqnarray}
where $\alpha$ is the viscous parameter.
In addition, the mass-loss rate $J$
per unit surface area becomes
\begin{equation}
   J = -2\dot{\rho}H = \frac{1}{2\pi r}\frac{d\dot{M}}{dr}
           =  \frac{\dot{M}_{\rm input}}{2\pi r_{\rm cr}}
                \frac{1}{r}.
\end{equation}

In the present non-dimensional unit
in terms of $c$, $r_{\rm g}$, and the Eddington luminosity $L_{\rm E}$,
these physical quantities are expressed as
\begin{eqnarray}
   \tau_0 &=& \frac{16\sqrt{6}}{\alpha} 
            \sqrt{{r}},
\\
   F_{\rm s} &=& \frac{3}{4}\sqrt{c_3}
                 \frac{1}{r^2},
\\
   P_0 &=& \sqrt{c_3}\tau \frac{1}{r^2},
\\
   J &=&    \frac{1}{r},
\end{eqnarray}
where the symbol ``hat'' (say, $\hat{r}$) is dropped.

Using these relations,
we can solve the basic equations, and obtain numerical solutions
at each radius $r$.
The example in the case of $\dot{m}=100$ and $\alpha=1$
is shown in figures 3 and 4.

\begin{figure}
  \begin{center}
  \FigureFile(80mm,80mm){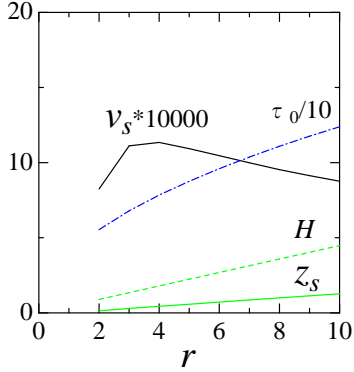}
  \end{center}
\caption{
Several quantities for each radius $r$:
The disk height $H$ (dashed curve) and 
the optical depth $\tau$ (chain-dotted one)
are from the critical model, while
the height $z_{\rm s}$ and velocity $v_{\rm s}$
(solid curves) at the flow top are the results of 
the present numerical calculations.
The quantities are normalized in units of $c$ and $r_{\rm g}$.
The parameters of the critical disk is
$\dot{m}=100$ and $\alpha=1$.
}
\end{figure}

\begin{figure}
  \begin{center}
  \FigureFile(80mm,80mm){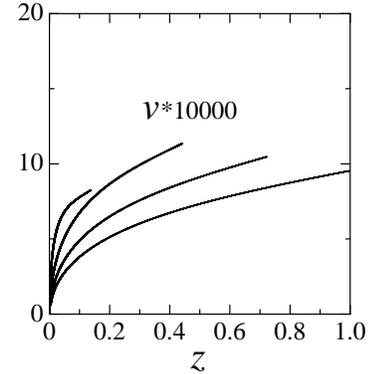}
  \end{center}
\caption{
Flow velocity $v$ (solid curves)
and optical depth $\tau$ (dashed ones),
as a function of height $z$
for several values of $r$.
The values of $r$ are, from left to right,
2, 3, 4, 5, 6, 7, and 8.
The quantities are normalized in units of $c$ and $r_{\rm g}$.
The parameters of the critical disk is
$\dot{m}=100$ and $\alpha=1$.
}
\end{figure}

In figure 3
we show several quantities for each radius $r$:
The disk height $H$ (dashed curve) and 
the optical depth $\tau_0$ (chain-dotted one)
are from the critical accretion disk model, while
the height $z_{\rm s}$ and velocity $v_{\rm s}$
(solid curves) at the flow top are the results of numerical calculations.
The quantities are normalized in units of $c$ and $r_{\rm g}$.

As can be seen in figure 3,
the flow height $z_{\rm s}$, where the optical depth is zero,
is about $1/4$ of the disk scale height $H$.
Moreover, the final speed $v_{\rm s}$ at the flow top
is quite small in this example; $v_{\rm s} \sim 0.001$.
It is noted, however, this speed is {\it not}
the final wind speed from the disk,
but the initial {\it starting} speed
of the mass loss from the disk surface.
With this starting speed,
mass loss from the disk surface takes place,
and is further accelerated in an optically thin regime
by the intense disk radiation fields.
It should be noted that
the final speed $v_{\rm s}$ gradually increases with decreasing $r$
since the flux $F_{\rm s}$ increases with decreasing $r$.
In the innermost region, however,
it decreases due to the relativistic gravity 
of the pseudo-Newtonian potential.

In figure 4
we show
the flow velocity $v$ (solid curves)
and optical depth $\tau$ (dashed ones),
as a function of height $z$
for several values of $r$.
The quantities are normalized in units of $c$ and $r_{\rm g}$.

As can be seen in figure 4,
the flow velocity $v$ roughly varies as
\begin{equation}
   v \propto z^{1/2}.
\end{equation}
This behavior is easily understood,
since equation (\ref{v1}) becomes
\begin{equation}
   v\frac{dv}{dz} = \frac{\kappa_{\rm abs}+\kappa_{\rm sca}}{c}F
\label{v1_NR}
\end{equation}
in the non-relativistic limit without gravity,
and easily integrated to give
\begin{equation}
   \frac{1}{2}v^2 = \frac{\kappa_{\rm abs}+\kappa_{\rm sca}}{c}Fz,
\label{v1_NR2}
\end{equation}
when the flux is constant.

\section{Concluding Remarks}

In this paper 
we have examined the radiative transfer flow in a luminous disk
in the subrelativistic regime of $(v/c)^1$,
while taking account of gravity of the central object.
The flow is assumed to be vertical, and
the gas pressure is ignored for simplicity,
while internal heating is assumed to be proportional to the density.
The basic equations are numerically solved
as a function of the optical depth $\tau$,
and the flow velocity $v$, the height $z$, the radiative flux $F$, and
the radiation pressure $P$ are obtained
for a given radius $r$, the initial optical depth $\tau_0$,
and the initial coditions at the flow base (disk ``inside''),
whereas the mass-loss rate $J$ is determined as an eigenvalue
of the boundary condition at the flow top (disk ``surface'').
For sufficiently luminous cases,
the flow resembles the case without gravity.
For less-luminous cases, however,
the flow velocity decreases, and
flow would be impossible due to the existence of gravity.
%

The radiative flow investigaed in the present paper
must be a quite {\it fundamental problem} for
accretion-disk physics and astrophysical jet formation,
although the present paper is only the second step (cf. Fukue 2005b), and
there are many simplifications at the present stage.

For example, we have ignored the gas pressure.
In general cases, where the gas pressure is considered,
there usually appears sonic points (e.g., Fukue 2002),
and the flow is accelerated from subsonic to supersonic.
In this paper we consider a purely vertical flow,
and the cross section of the flow is constant.
If the cross section of the flow increases along the flow,
the flow properties such as a transonic nature would be influenced.

Fully relativistic effects are also important.
In such a fully relativistic case,
where the flow speed is of the order of the speed of light,
the boundary condition at the flow top should be carefully treated
(cf. Fukue 2005c).
The radiation drag becomes much more important.

There remain many problems to be solved.

\vspace*{1pc}

This work has been supported in part
by a Grant-in-Aid for the Scientific Research (15540235 J.F.) 
of the Ministry of Education, Culture, Sports, Science and Technology.


\end{document}